\documentclass[preprint]{aastex}
\usepackage{graphicx}

\begin{document}

\title{On the Interpolation of Model Atmospheres and High-Resolution Synthetic Stellar Spectra}

\author{
Sz.~M{\'e}sz{\'a}ros\altaffilmark{1,2}, 
C.~Allende~Prieto\altaffilmark{1,2}
}

\altaffiltext{1}{Instituto de Astrof{\'{\i}}sica de Canarias (IAC), E-38200 La Laguna, Tenerife, Spain}
\altaffiltext{2}{Departamento de Astrof{\'{\i}}sica, Universidad de La Laguna (ULL), E-38206 la Laguna, Tenerife, 
Spain}

\begin{abstract}
We present tests carried out on optical and infrared stellar spectra 
to evaluate the accuracy of different types of interpolation.
Both model atmospheres and continuum normalized fluxes were interpolated. 
In the first case we used linear interpolation, and 
in the second linear, cubic spline, cubic-Bezier and quadratic-Bezier methods. 
We generated 400 ATLAS9 model
atmospheres with random values of the atmospheric parameters 
for these tests, spanning between $-2.5$ and $+0.5$ in [Fe/H], 
from 4500 to 6250 K in effective temperature, and 1.5 to 4.5 dex
in surface gravity. 
Synthesized spectra were created from these model atmospheres, and compared with spectra derived by interpolation. 
We found that the most accurate interpolation algorithm among those considered in flux space is cubic-Bezier, 
closely followed by quadratic-Bezier and cubic splines. 
Linear interpolation of model atmospheres results in errors about a factor of two larger than 
linear interpolation of fluxes, and about a factor of four larger than high order flux interpolations. 

\end{abstract}

\section{Introduction}

Even within the framework of classical LTE 1D models, the
calculation of a stellar model atmosphere takes a finite amount
of time, which can be significant when complex opacities are
involved. Massive spectroscopic surveys require large numbers
of models spanning a wide range of parameters and can become
very time consuming. Depending on the algorithm used, 
even the analysis of a single star may require many model 
atmospheres to evaluate the performance of different
combinations of parameters.  

In practice, the need for model spectra for many parameter 
combinations is satisfied by taking one or several shortcuts
that avoids the actual calculation of self-consistent models.
The most wide-used strategy is some sort of interpolation,
either in the model atmosphere (the run with height of the
main thermodynamical variables), or in the emerging radiative 
fluxes or intensities. Different recipes have been used, and
codes circulate among researchers, but few have been published
and thoroughly tested. 

In this paper we perform a battery of tests in order to quantify
the typical errors incurred when interpolating model atmospheres
or the model fluxes calculated from them. Section 2 describes
our calculations and \S 3 our results, with a summary provided 
in \S 4.

\section{Calculations}

We generated a regular grid of ATLAS9 model atmospheres 
with [Fe/H] from $-2.5$ to $+0.5$ in steps of 0.5 dex,
$T_{\rm eff}$ from 4500~K to 6250~K in steps of 250 K, and log~g from 1.5 
to 4.5 dex in steps of 0.5 dex. We also calculated 400 additional models with 
random parameters within the boundaries of
the regular grid. The microturbulence was chosen to be constant at 
2 km s$^{-1}$ in all our calculations. 

The relatively small range of $T_{\rm eff}$ is to ensure that all ATLAS9 models are fully 
converged throughout the entire atmosphere. Since ATLAS9 start to experience 
convergence problems in the outer layers of the atmosphere below 4250~K and above logg = 4, 
especially at low metallicities, we chose to omit that region from the calculations. Stars warmer 
than 6250~K have spectra dominated mainly by hydrogen lines, thus it is expected that interpolation 
errors will decrease as temperatures increase, and the examples of spectra with 6250~
K are good representation of warmer stars.
After careful consideration we decided that the above values gave the largest 
range in all three parameters combined. 

Two representative  wavelength regions were chosen, 
one in the optical, and including both weak and strong spectral lines
(the Mg I$b$ triplet) between 516.5 and 519.5 nm, 
and one in the near-infrared, and in particular the H-band window, 
targeted by the Apache Point Galactic Evolution Experiment 
\citep[APOGEE,][]{allende01,eisenstein01,wilson01}, spanning between 
1509.1 and 1699.5 nm. 

We carried out two types of tests, one involving interpolation
in the model atmospheres directly, and a second one involving interpolation in the emergent fluxes. 
In the first type of test, we interpolated models with the chosen random parameters from the 
grid, and synthesized the emergent spectra. 
The comparison was made between these and the fluxes 
calculated from the models with the randomly generated parameters.

In the second case, we calculated the emergent spectra
for the models generated from the random parameters (true flux), and the evenly 
distributed models. Then, we 
interpolated the spectra from the evenly spaced models to the sets of random parameters and examined the 
differences in continuum normalized flux relative to the true flux. 
The flux interpolation tests were performed with linear (F(L)), cubic spline (F(CS)), cubic-Bezier (F(CB)), and 
quadratic-Bezier (F(QB)) methods, while in the case of the model atmosphere interpolations (MA) we only 
explored the linear algorithm. 

\subsection{Model atmosphere calculations}

Our regularly spaced ATLAS9 model atmospheres were 
calculated  as described by \citet{meszaros01}. However, there are some 
differences between these models  and those presented by 
\citet{meszaros01}\footnote{http://www.iac.es/proyecto/ATLAS-APOGEE/}:
those used here correspond to different versions of the line data,
and an older version of the code, plus a number of other differences
regarding the configuration of the input for the ATLAS9 code. 
The calculations used here are older, and the updated ones are
to be preferred, but since these details are irrelevant for
the evaluation of the interpolation accuracy, we have
chosen to retain the custom-made calculations. 

The opacity distribution functions (ODFs) and
Rosseland opacities needed as input
were calculated using the DFSYNTHE and KAPPA9 codes, while the model atmospheres
were generated with the linux version of the
ATLAS9 code \citep{kurucz05, kurucz01, sbordone02, sbordone01}. The ODF calculations
followed the method described by \citet{castelli01, castelli02}, while the model atmosphere calculations 
are detailed in \citet{meszaros01}. ATLAS9 gives excellent convergence in the
parameter range chosen above. 
  
In addition to the regular grid, we made calculations for 
400 additional random models. The parameters for these were drawn from random 
uniform distributions across the chosen ranges. 
These calculations are  fully consistent 
with the models in the grid. 

\subsection{Model atmosphere interpolation}

We interpolated  model  atmospheres for the 400 sets of 
random parameters. For each target model, we identified
the 8 immediate neighbors with higher  and lower values 
for each parameter in the grid, calculated by numerical
integration the Rosseland optical depth for each, 
re-sampled all the thermodynamical 
quantities in the atmosphere (temperature, gas pressure, 
and electron density) on a common optical depth scale for
all models by linear interpolation, and 
then interpolated, linearly, all the thermodynamical 
quantities to the parameters ($T_{\rm eff}$, $\log g$, 
and [Fe/H]) of the target model. 

Other quantities included
in the models (Rosseland opacities,
radiative pressure, etc.) were also interpolated in
the same way. The interpolations were carried out using the {\tt kmod}
code. This code has already been used in a number of investigations
\citep{reddy01, reddy02, yong01}. It
is written in IDL and it is publicly 
available\footnote{http://leda.as.utexas.edu/stools}.

\subsection{Calculation of model fluxes}
 
We calculated model fluxes for all model atmospheres using 
the ASS$\epsilon$T spectral synthesis code \citep{koesterke01, koesterke02}
with detailed continuum opacities by \citet{allende02} and updates from \citet{allende03}. 
Line data come mainly from the calculations and compilations by Kurucz
(available from his website\footnote{http://kurucz.harvard.edu/}), enhanced with damping constants from 
\citet{barklem01} when available.

We adopted 
solar reference abundances as in \citet{asplund01}, and
the compositions used in the calculations of  the model 
atmospheres were consistent
with those adopted in the spectral synthesis. We underline,
however, that the interpolation tests performed here
are fairly insensitive to the particular choices for
the reference solar composition and the atomic and molecular 
data, as long as these choices are reasonable and lead
to spectra that resemble approximately the modeled stars.
The most critical aspect, is to ensure that the calculations
for the models in the grid and those with random parameters
used for tests, are completely consistent.  

\subsection{Flux Interpolations}

Flux interpolations are consecutively performed in surface
gravity, effective temperature, and metallicity.  Quadratic
and cubic Bezier interpolation are implemented with control
values that make the algorithm identical to Hermite interpolation,
and therefore both the interpolating function and its derivatives are
continuous. In the case of cubic Bezier
interpolation, quadratic Bezier interpolation was forced for those dimensions
for which the target parameters  where in the intervals adjacent
to the edges of the grid.

The Bezier interpolations were implemented afresh in the FORTRAN
spectral fitting code {\tt FERRE} \citep{allende04} following
the description  by \citet{auer01} and references therein. Cubic
splines interpolation was included calling a subroutine
from the library provided with the book by \citet{chapman01},
which follows the discussion given in \citet{press01}.

\section{Discussion}

Before examining the errors in the fluxes, we compared interpolated model atmospheres with the calculated ones.
The linear interpolation does a good job at estimating the real atmospheric structure; 
the temperature differences were between $2-3\%$, the pressure and electron density differences were between
$1-2\%$. This translates to 10$-$30~K average differences in temperature in case of cool atmospheres above 
log$~\tau_{Ross}<1$, from where all of the spectral lines form. 
This difference slightly increases as the effective temperature gets higher. 

To measure the amount of error each interpolation method makes, 
we calculated the same statistics for each case. In the case of absolute fluxes, we only examined the average 
differences throughout the whole wavelength range, to track how much
the continuum level changes. In the case of optical 
spectra, the atmosphere interpolation gives 1$-$5\% smaller continuum
levels than the direct calculations, and we find that 
the linear flux interpolation 
gives -0.3$- +$0.5\% relative differences, usually shifted slightly higher than the 
true continuum in case of the optical spectrum, but not in the infrared.  
The errors are smaller in the IR region, 0.5$- +$1\% in the continuum 
level when atmospheres are interpolated and 0.01$-$0.1\% with linear 
interpolation in flux. 

Two examples of the differences between  
interpolated and direct calculations of continuum normalized 
spectra are given in Figures 1 and 2. Figure 1 shows a metal-poor
G-type giant, while Figure 2 shows a metal rich, K-type dwarf --  
these two cases represent the extremes of our
parameters space. 
The metal-poor spectra only show large differences in the line cores. 
The average error can be up to 
1\% in the optical and infrared metal-poor warm spectrum, 
while in case of the metal-rich cool spectrum this goes up to
5$-$10\% in the optical and 1$-$2\% in the infrared. 
The differences in the line profile for the cooler atmosphere increase by a factor of 4 in both 
wavelength regions compared to the metal-poor case due to 
weak atomic metal and molecular lines very sensitive to small 
temperature changes in the atmosphere. In the infrared spectra of the G-tpe metal-poor star, the 
differences are dominated by the hydrogen Brackett lines, 
while in the optical these are associated with the 
magnesium triplet lines. This 
picture changes dramatically in case of the metal-rich, cool dwarf 
(Figure 2), where the Brackett lines disappear and 
the "noise" increases greatly in the continuum of the 
IR spectrum due to weak atomic and molecular lines (CN, OH, and CO). Also, 
relative errors in the line cores are significantly higher than near the continuum. 

To measure the overall performance of the interpolation 
methods across the parameter range, the differences respect to the true fluxes were calculated at each 
wavelength in the continuum normalized spectra. The average, 
standard deviation and maximum deviation 
above and below the continuum in the whole wavelength 
range were determined. The last two of these three parameters track different 
aspects of the error distribution. 
The standard deviation gives an estimate 
of the overall changes in the full flux range, while 
the maximum deviation shows mainly the differences in the line cores, where errors tend to be largest. 
The differences vary depending on where one 
samples the spectrum: they are small in the continuum, 
but typically larger in the line cores. 

Figure 3 illustrates how the average differences 
depend on metallicity, effective temperature and gravity. The errors, both the average  
errors and the standard deviation, 
grow with increasing metallicity and decreasing temperature, 
but they do not depend so much on surface gravity. These 
correlations are easily understood as a result of an 
increased number of molecular and atomic absorption lines appearing in 
the spectrum at high metallicity and low temperature. 

Both in the model atmosphere and linear flux interpolation methods, it is clearly visible that
errors are significantly enhanced for values of the metallicity half way 
between the nodes of the grid. 
The evenly distributed models spanned $-2.5$ to 
$+0.5$ in [Fe/H] with steps 0.5, and in the vicinity of these 
values the linear interpolation gives significantly lower errors. 
Using higher order interpolations makes this effect to disappear.
None of the methods is sensitive to this issue along the axis for
effective temperature (nodes spaced every 
250~K) or log g (nodes every 0.5 dex). 

The overall results for each test are listed in Table 1 and illustrated  
in Figure 4. In the optical, the average differences are about 0.17$\pm$0.19\% 
for the case of MA interpolation, and smaller than 0.07\% 
for all the interpolations in flux. The standard deviation of the relative differences 
paints a more dramatic picture. The MA method clearly shows the 
largest deviation up to 1\% near the continuum level 
in the optical and, while F(CB) is the most accurate method with only 0.1\% errors. 
In the infrared all interpolation 
methods perform well giving much smaller errors than in the optical, close to an average of 
0$-$0.02\% differences. The scatter is about four times larger in the case of MA 
than in any other methods in both optical and infrared. 

Figure 5 shows the maximum deviation of differences 
below and above the continuum level in the case of the 
cubic-Bezier interpolation. At optical wavelengths, the
 differences are generally higher than in the infrared, 
but they are usually smaller than 0.01 in $\Delta$~F (1$-$2\%). 
While a reduction of the errors near the grid nodes was 
not visible in the average differences, here it is apparent
 as a function of metallicity below [Fe/H]$=-1$, and the effect 
increases as [Fe/H] decreases. The interpolation gives the 
highest errors in the line cores, i.e. in the highest atmospheric layers
the spectrum is sensitive to.

\section{Conclusions}

We conclude that if model spectra need to be interpolated, 
the best way to proceed, among those explored, is to
use high-order interpolation in continuum-normalized fluxes. 
Linear interpolation of model atmospheres leads to 
about a factor of 3$-$5 higher error than high order interpolations in flux, 
at about $-$0.17$\pm$0.19\% average differences in the optical, and 
0.0004$\pm$0.038\% in the infrared. Linear interpolation in flux space gives a factor of two smaller
errors than using model atmospheres, but there is still a factor of two improvments found using high order functions. 
The most accurate flux interpolation method is cubic-Bezier with average errors 
of 0.011$\pm$0.047\% in the optical and 0.0011$\pm$0.0075\% in the infrared for our 
grids and random parameters. The cubic spline and quadratic-Bezier 
interpolations lead to only marginally higher errors than this. We conclude that 
interpolation errors become visible at $S/N = $10$-$20 in the line cores in case of 
model atmosphere interpolations in the optical, and at $S/N = $20$-$40 in the infrared, 
while the errors from cubic-Bezier flux interpolation will only show up above 
$S/N \sim $100$-$200 in both wavelength ranges. 

We must stress that only one code for the interpolation of
model structures has been tested. Other codes with somewhat different strategies could provide
somewhat different performances. We have 
allowed our high-order interpolations in flux to 
get to any value, even if artificial extrema are created \citep[see the discussion by][]{auer01}. 

Our tests are restricted to stars with spectral types
roughly between F5 and K5. Nevertheless, we expect that the 
tendencies seen in our study will persist outside this domain.
For example, we expect that errors will continue to reduce
for the interpolation of stars with warmer temperatures, and
to increase at temperatures under 4500~K. 

Our analysis includes only linear and polynomial interpolation,
but more sophisticated schemes are available and their 
performance could be somewhat different. Further work 
dealing with principal component analysis (PCA), neural
networks, and other algorithms would be a natural followup
of this investigation. 

\acknowledgements{We would like to thank our collegues at the SDSS-3 APOGEE survey for providing continuous discussions
on various interpolation techniques. We greatly appreciate the contribution of Lars Koesterke on spectral 
synthesis with the ASS$\epsilon$T code. We are also grateful for the referee for his helpful suggestions.}

\clearpage

\begin{figure}
\includegraphics[width=4.6in,angle=270]{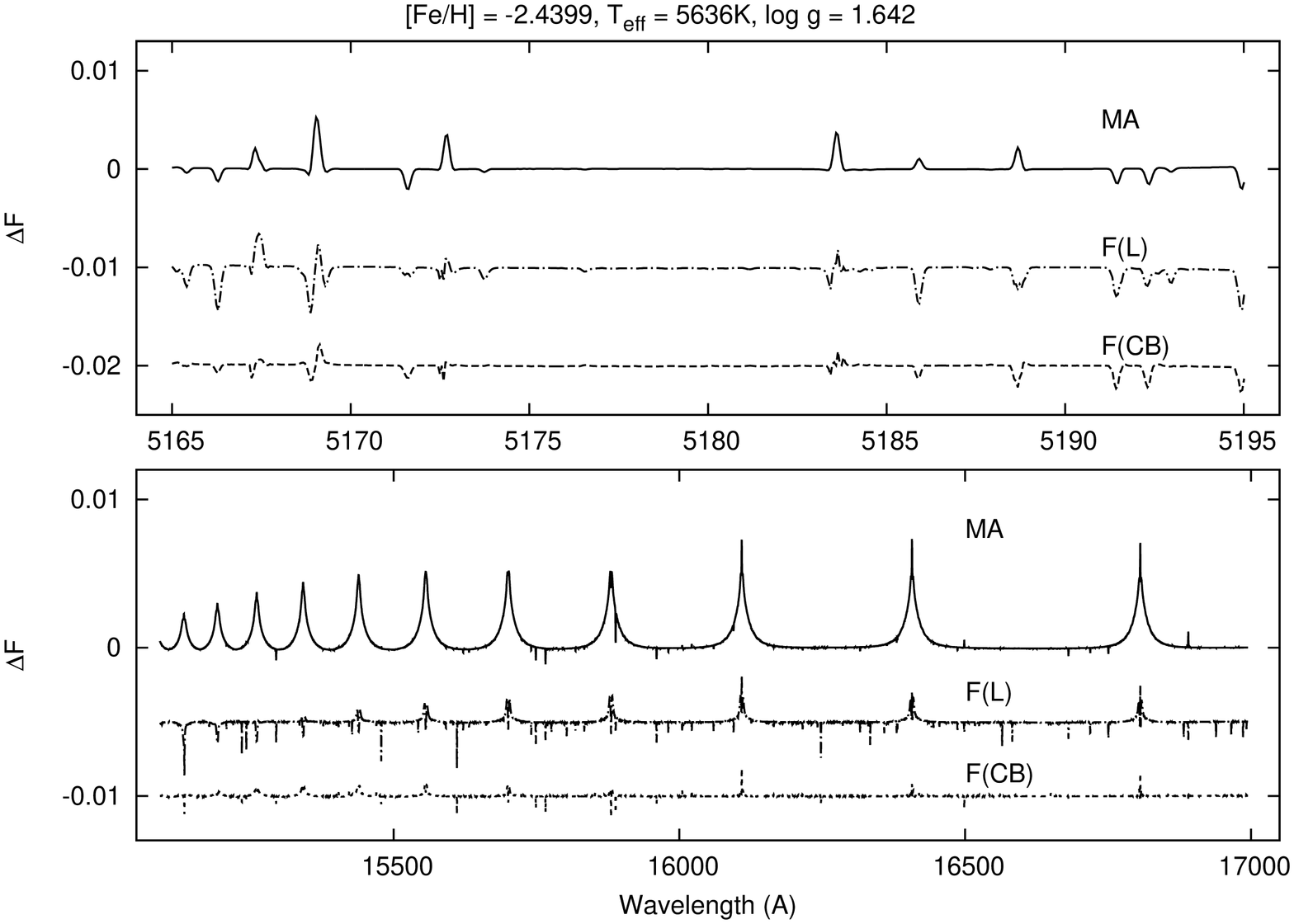}
\caption{Examples of differences of continuum normalized spectra in case of a metal-poor hot giant. On the left panels
relative differences, while on the right side normal differences compared to the true flux are shown. Only three 
interpolation methods are presented:
model atmosphere (MA), linear flux F(L), and cubic-Bezier flux F(CB). F(L) and F(CB) are shifted down by 0.01 and 0.02
respectively in the optical, and by 0.005 and 0.01 in the infrared spectrum to aid visibility. The MA method gives 
the largest errors, while F(CB) shows the smallest differences.}
\end{figure}

\begin{figure}
\includegraphics[width=4.6in,angle=270]{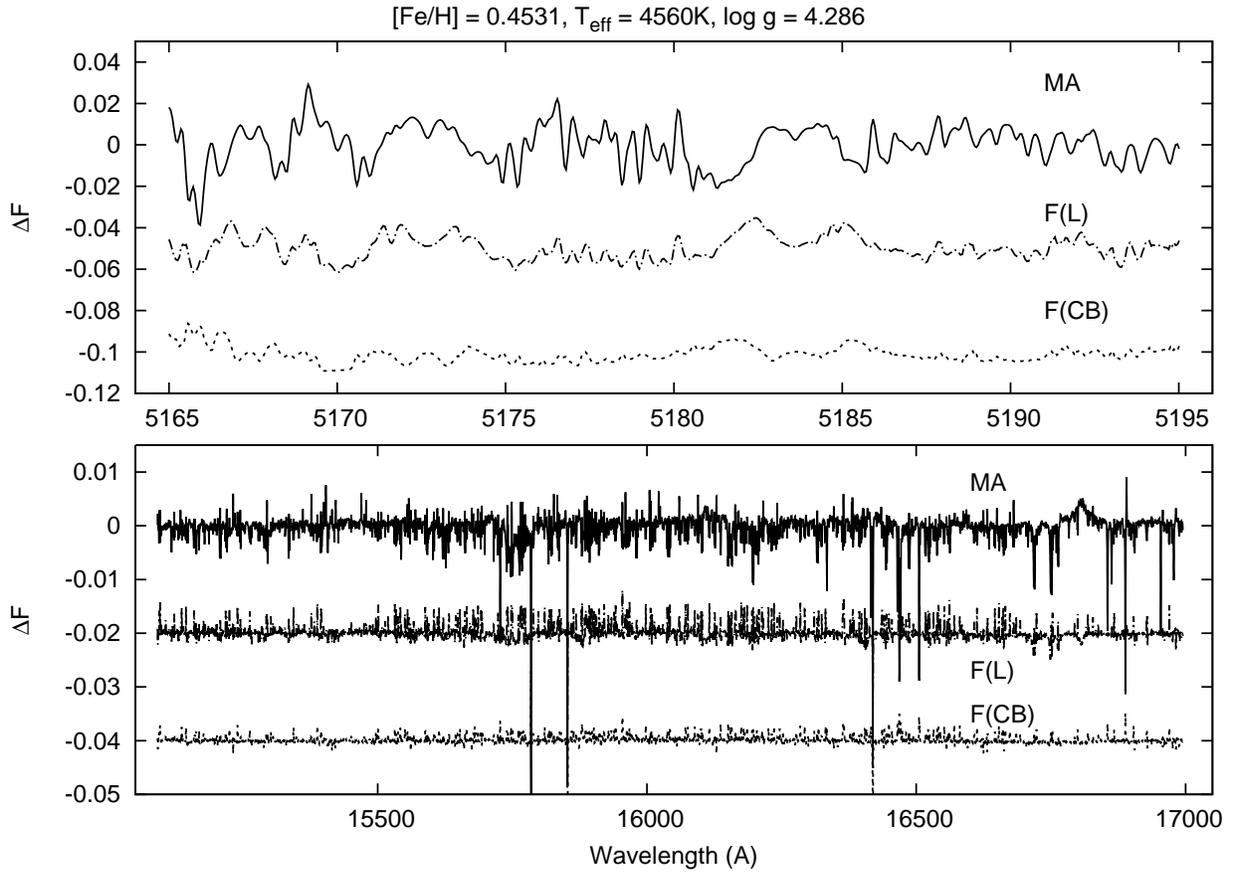}
\caption{Examples of differences of continuum normalized spectra in case of a metal-rich cool dwarf. F(L) and 
F(CB) are shifted down by 0.05 and 0.1 
respectively in the optical, and by 0.02 and 0.04 in the infrared spectrum to aid visibility.}
\end{figure}

\begin{figure}
\includegraphics[width=4.6in,angle=270]{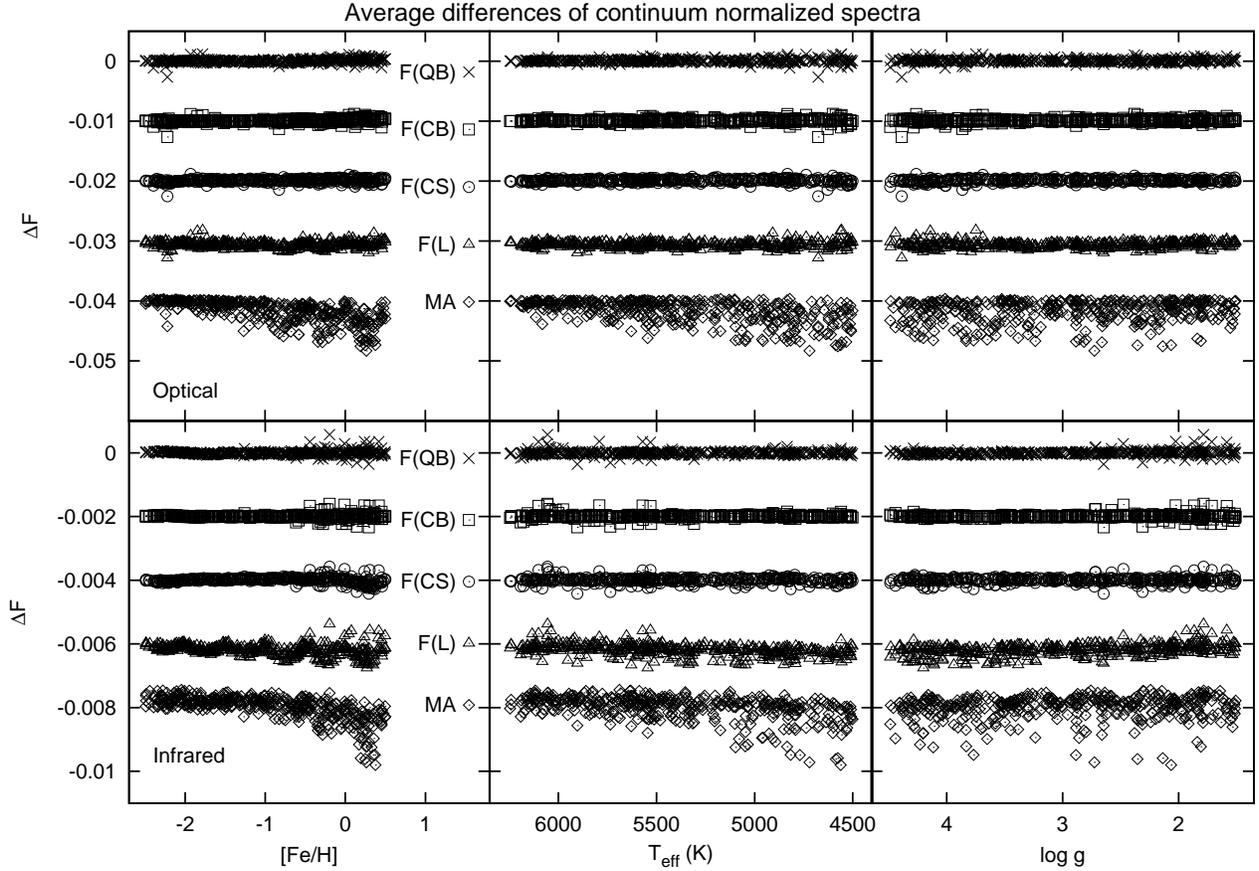}
\caption{Average differences through the entire spectrum as a function of [Fe/H], T$_{\rm eff}$, and log g. Annotations
are explained in Section 2. Results for each interpolation method are shifted by 0.01, and 0.002 in
optical and infrared respectively. The linear
model atmosphere (MA) interpolation give significantly higher errors compared to the flux interpolations. Errors from all
methods increase with increasing metallicity and decreasing temperature, however none depends on the gravity.}
\end{figure}

\begin{figure}
\includegraphics[width=4.6in,angle=270]{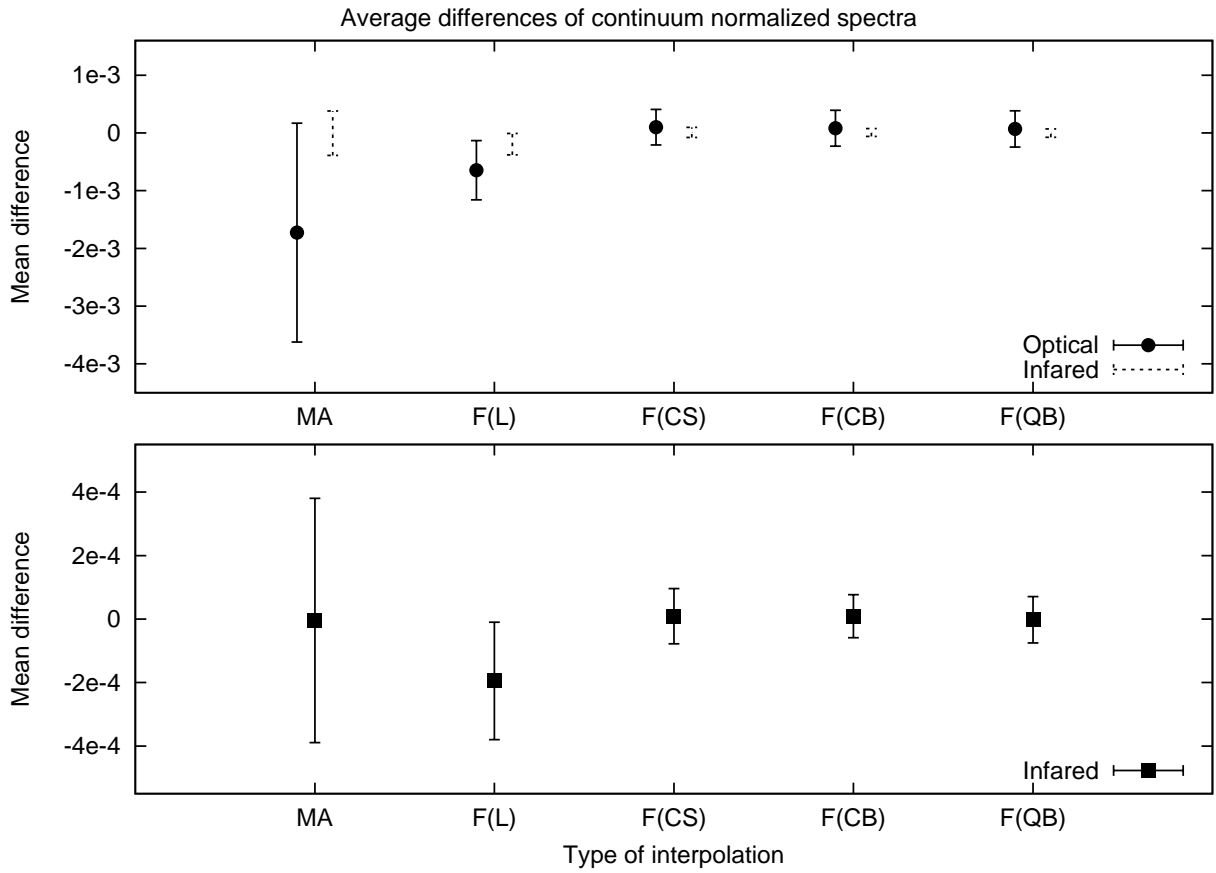}
\caption{The average and standard deviation of the average relative differences of all 400 test atmosphere in 
continuum normalized flux for each interpolation method. Annotations are explained in Section 2. The linear
model atmosphere (MA) interpolation show about a factor of 3 higher scatter than the flux interpolations, while the
cubic-Bezier method gives the smoothest results.}
\end{figure}

\begin{figure}
\includegraphics[width=4.6in,angle=270]{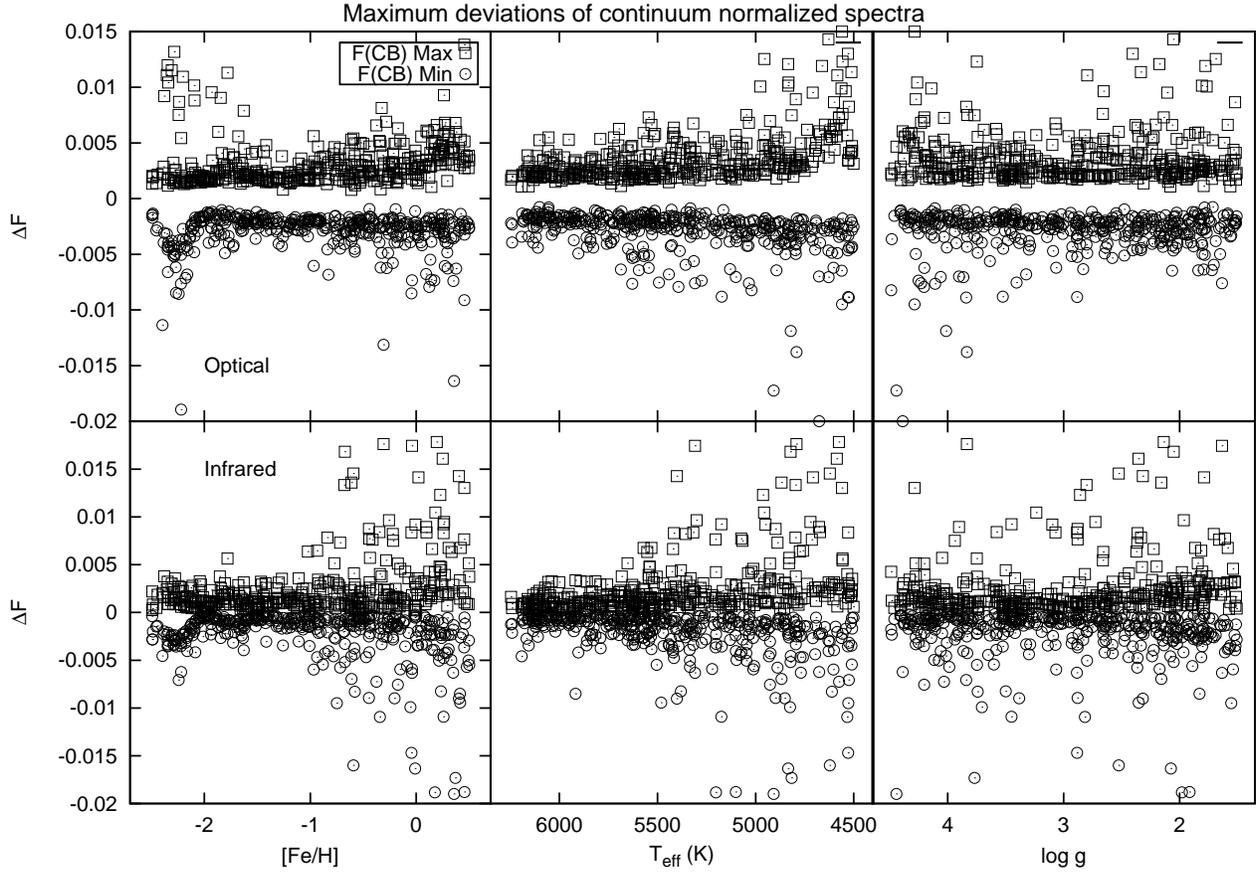}
\caption{The relative maximum deviation above and below the continuum level, which tracks the largest differences in the
line core profiles. Some outliers in the IR plots were not plotted to show the dependence on the physical parameters
better. }
\end{figure}

\clearpage

\begin{deluxetable}{lrrrrr}
\tabletypesize{\scriptsize}
\tablecaption{Overall statistics of all the interpolation methods}
\tablewidth{0pt}
\tablehead{\colhead{$\Delta F$} & \colhead{MA} & \colhead{F(L)} & \colhead{F(CS)} & \colhead{F(CB)} & \colhead{F(QB)}}
\startdata
Average (OP) & -1.727e-3 & -6.458e-4 & 1.004e-4 & 8.263e-5 & 6.976e-5 \\
S.D. (OP) & 1.895e-3 & 5.132e-4 & 3.081e-4 & 3.103e-4  & 3.140e-4 \\
Min (OP) & -0.058 & -0.021 & -0.018 & -0.019 & -0.021 \\
Max (OP) & 0.029 & 0.027 & 0.012 & 0.014 & 0.014 \\
\hline
Average (IR) & -4.274e-6 & -1.950e-4 & 8.928e-6 & 9.124e-6 & -2.230e-6 \\
S.D. (IR) & 3.849e-4 & 1.850e-4 & 8.711e-5 & 6.775e-5 & 7.295e-5 \\
Min (IR) & -0.195 & -0.104 & -0.108 & -0.107 & -0.105 \\
Max (IR) & 0.0109 & 0.095 & 0.096 & 0.097 & 0.097 \\
\enddata
\tablecomments{OP: optical, 516.5$-$519.5 nm, IR: infrared, 1500$-$1700 nm. The type of interpolations are 
the following: MA: model atmosphere, F(L): linear flux, 
F(CS): cubic spline flux, F(CB): cubic-Bezier flux, F(QB): quadratic-Bezier flux}
\end{deluxetable}

\end{document}